\def\bq{\begin{equation}}
\def\eq{\end{equation}}
\def\bqa{\begin{eqnarray}}
\def\eqa{\end{eqnarray}}
\def\bqb{\begin{eqnarray*}}
\def\eqb{\end{eqnarray*}}
\documentstyle[12pt]{article}\textwidth 16cm
\def\pr#1#2#3{ Phys. Rev. ${\bf{#1}}$ (#2) #3 }
\def\prl#1#2#3{ Phys. Rev. Lett. ${\bf{#1}}$ (#2) #3 }
\def\pl#1#2#3{ Phys. Lett. ${\bf{#1}}$ (#2) #3 }

\def\np#1#2#3{ Nucl. Phys. ${\bf{#1}}$ (#2) #3 }
\def\zp#1#2#3{ Z. Phys. ${\bf{#1}}$ (#2) #3 }

\def\ie{{\it i.e.\/}}
\def\eg{{\it e.g.\/}}

\def\etal{{\it et.al.\/}}

\global\nulldelimiterspace = 0pt

\def\roughly#1{\mathrel{\raise.3ex
    \hbox{$#1$\kern-.75em\lower1ex\hbox{$\sim$}}}}
\def\lsim{\roughly<}

\def\ol#1{\overline{#1}}

\def\L{ {\cal L }}
\def\O{ {\cal O }}
\def\sw{s_W}
\def\cw{c_W}
\def\swd{s^2_W}

\begin{document}
\pagenumbering{arabic}
\thispagestyle{empty}
\def\thefootnote{\fnsymbol{footnote}}
\setcounter{footnote}{1}

\begin{flushright} PM/94-44 \\ THES-TP 95/01 \\
January 1995 \\
 \end{flushright}
\vspace{2cm}
\begin{center}
{\Large\bf Residual heavy quark and boson
interactions: the role of the $Zb\bar b$ vertex}
\footnote{Partially supported by the EC contract CHRX-CT94-0579.}
 \vspace{1.5cm}  \\
{\large G.J. Gounaris$^a$, F.M. Renard$^b$ and C.
Verzegnassi$^c$}
\vspace {0.5cm}  \\

$^a$Department of Theoretical Physics, University of Thessaloniki,\\
Gr-54006, Thessaloniki, Greece.\\
\vspace{0.2cm}
$^b$ Physique
Math\'{e}matique et Th\'{e}orique,
CNRS-URA 768\\
Universit\'{e} Montpellier II,
 F-34095 Montpellier Cedex 5.\\
\vspace{0.2cm}
$^c$ Dipartimento di Fisica,
Universit\`{a} di Lecce \\
CP193 Via Arnesano, I-73100 Lecce, \\
and INFN, Sezione di Lecce, Italy.\\

\vspace {1cm}

{\bf Abstract}
\end{center}
\noindent
We establish the most general parametrization of the new physics
tested by present precision measurements and possibly responsible
for any deviation of
the $Z\to b \bar b $ amplitude from its Standard Model result,
under the assumption that it is CP symmetric and is induced by degrees
of freedom which are
too heavy to be ~directly produced at the future Colliders.
This is achieved by writing the complete list of the
$SU(3)_c\times SU(2) \times U(1) $ gauge
invariant and CP symmetric $dim=6$ operators, involving only
quarks of the third family and/or bosons.
The quark ~containing operators are divided into two classes,
according to whether they involve the $t_R$ field or
not. Each class contains 14 quark operators. We then
proceed to derive the constraints from present precision
measurements, on the first class of the 14 $t_R$ involving
operators. We show that the $Zb \bar b$ vertex plays a fundamental role
to discriminate not only between these operators, but also between this whole
scheme  and an alternative one like \eg\@ an MSSM description with light
chargino and neutral Higgs.\\
\vspace{0.5cm}
\def\thefootnote{\arabic{footnote}}
\setcounter{footnote}{0}
\clearpage

\section{Introduction}
A complete and rigorous investigation of the status of the Standard
Model (SM) requires a critical analysis of its various sectors. As of
today, this has been possible only for the fermionic sector,
thanks to the impressive experiments that have been performed
at LEP1, at SLC and at
lower energy \cite{lep1,lep1b,slc}. On the contrary
the status
of the bosonic sector (gauge and scalar boson interactions) is not yet
empirically established to a convincing precision.
Although a number of indirect tests concerning \eg\@ the triple
gauge couplings already indicate that, there also, the deviation
from SM cannot be dramatic \cite{BMT,Hag,DeR}, it
is generally felt that more ~accurate tests at higher energy colliders
are required in order to be able to state that we have really
tested the theory.\par

As far as the fermionic sector is concerned, it is certainly
true that the agreement with the SM predictions is amazingly good (up
to a few permille) in the \underline{light} fermionic part. The
situation is slightly less triumphant in the heavy quark
sector, where, as it has been exhaustively discussed in previous
papers \cite{lep1b}, the experimental value of $\Gamma_b$
(the Z width into $b\bar b$), shows a small discrepancy from the SM
prediction, which  increases with the
top mass $m_t$ and reaches the $2\sigma$ level for $m_t$ values
in the region of $175 GeV$ \cite{topfnal}. In addition to this, the top
quark interactions themselves are also to a large extent empirically
unknown.\par

Within the SM, the most important difference
between the $Zb\bar b$ vertex involved in the
possibly "rebel" $Z\to b\bar b$ width, and the light $Zf\bar f$
vertices, arises at the one loop level and has the form of
a contribution
proportional to $m_t^2$. Such a contribution appears in the
$Zb\bar b$ vertex only and  originates, in a $R_{\xi}$ gauge,
from the Yukawa coupling of a charged would-be Goldstone boson
with a $(t\bar b)$ pair.
Since the corresponding contribution to light fermion vertices
is negligible, one suspects that a kind of link should
somehow exist between the heavines of (one of) the  quarks of the
third family  and the possibility
that the SM predictions for this family are  "slightly" inadequate .
A similar inadequacy may apply to the bosonic sector
as well.\par

In this spirit, we subscribe to the feeling that the
fact that the $t$ quark and the  $(W,\ Z)$ pair are much heavier than
the leptons or the
other quarks is not casual, but rather deeply related
to the scalar sector of the theory, on whose origin it might perhaps
open one day an illuminating window.
Thus, a kind of NP may exist, originating from the scalar sector,
which could induce new interesting phenomena in the gauge boson,
Higgs and top interactions, and which may have already been
"seen", in the peculiarities of $Z\to b \bar b$.
As far as the $Z\to b\bar b$ decay
is concerned, this NP should appear in the form of contributions
enhanced by some power of the
heavy top quark mass. \par

One popular way of describing this kind of New Physics
(NP) is that of
assuming that it corresponds to an extension of the SM in which
all extra new degrees of freedom appear at a scale $\Lambda$
that is much heavier than the electroweak scale; \ie\@
$\Lambda \gg v$. At present energies, the effects of NP are
completely described by
integrating out all these new heavy degrees of freedom using
standard effective lagrangian techniques \cite{Georgi}. In this
approach one has ~thoroughly examined until now only the
possibility that this NP is entirely contained in the bosonic sector,
where it has been satisfactorily described in terms of
11 independent $dim=6$ gauge invariant operators \cite{Hag}. These
purely bosonic operators induce anomalous triple gauge boson
couplings at the tree level \cite{Hag}, and  at the 1-loop level
they also affect the fermionic vertices. In particular two of
these operators also
create at 1-loop \@ $m_t^2$ corrections to the $Z\to b \bar b$
amplitude, which could provide an explanation for
the possible deviation of $\Gamma_b$ from its SM value \cite{RV}.\par

With the exception of the very special case of the $Zt\bar t$
vertex ~considered in \cite{Peccei}, anomalous direct gauge boson-fermion
interactions, possibly involving also the Higgs particle, have been
disregarded up to now.  As stated above, the neglect of
anomalous gauge boson and
fermion interactions appears well motivated for light fermions.
It does not appear justified though, in cases where a
$t$ quark is participating, like $t$ and $b$ physics. A
fortiori, then, such anomalous interactions should
be investigated, ~particularly also since they can teach us
something about $Z\to b \bar b$.\par

The aim of this paper is that of
establishing a general description for
the residual NP interactions that may directly affect the quarks of the
third family. Assuming that the NP is CP symmetric and that it obeys
$SU(3)_c \times SU(2) \times U(1)$ gauge invariance, we
 classify all possible  $dim=6$ operators that could be induced
by it  at the present low energies. For purely bosonic operators,
this has already been done \cite{Hag}. Here we establish
the operators involving quarks of the third family only,
possibly together with gauge
and/or Higgs bosons. No light quarks
(from the first two families) or leptons are ~allowed.
The complete set of the purely bosonic and the above "third family"
operators should provide
a full description of NP for ~energies lower than the threshold for the
{}~excitation of the new degrees of ~freedom that may exist.
After this classification, we investigate what the existing
experimental information on $Zb\bar b$ can teach
us about these operators.  \par

Under the previous general assumptions, we find  28 independent
"third family" operators, which we
classify in two classes.
The first class contains 14 members which all involve the $t_R$
field. Since it is precisely the $\bar q_L t_R
\tilde \Phi$ combination which characterizes the top mass in SM,
it is natural to assume that the $t_R$ involving "top" operators
have a "strong ~affinity" to the scalar sector and, therefore, some of
them may get enhanced by it. Incidentally, a similar "strong
affinity" also applies to (some of) the 11
purely bosonic operators \cite{dyn}.
On the opposite side, currents  like \eg\@
$(\bar q_L \gamma^{\mu} q_L)$ have nothing to do with
the top quark mass. Consequently, the related operators are put in a
second class, as we feel that the possibility that they are
enhanced by NP is rather remote.\par

Therefore, we end up with a picture where NP is described in terms
of an Effective Lagrangian containing the 14 "top
operators" of the first class and the 11
{}~purely bosonic ones mentioned above. Since the ~consequences of
the purely "bosonic" operators have already been fully
analysed, we concentrate in this paper on the  14 $t_R$
involving ones.
These operators induce anomalous effects in direct processes
like \eg\@ top quark  production and
decay, and also indirect effects in processes where a
virtual top quark appears as intermediate state.\par

The analysis of direct processes will require a  clear and copious
production of top quarks ~which should be possible at future
colliders like \eg\@ LHC, NLC or maybe ~the Tevatron, after an
important development program. Since this is not
the most urgent point, we leave it for the future, and
we concentrate instead on the indirect processes for which
experimental results are presently available. We then find that
existing data can give
useful constraints on some of the "top-operators," and
provide an orientation on which operators one should retain in the
future analysis of the direct processes.\par

In Sect.2 we give the full list of the 28 CP symmetric,
$SU(3)_c \times SU(2) \times U(1)$ gauge ~invariant,
$dim=6$  "third family" operators of the first and second
class. For ~completeness, we also give the 11 "bosonic"
operators established in \cite{Hag}.
We  then derive the constraints that can be obtained from the
light fermionic sector ~using the LEP1,
SLC and low energy experiments. They are of two different types.
Firstly, those from
the  light fermionic processes (\ie\@ those not involving b
quarks), which are sensitive at 1-loop to top-operator
{}~contributions to the ~gauge boson self-energies.
Using these, we calculate in Sect.3 the effects on the relevant
$\epsilon_i$
parameters which establish constraints
on four independent top-operators.
Secondly, in Sect.4 we turn to the partial decay width
 $Z\to b \bar b$ and to the b asymmetry
\cite{CRV}, which provide constrains on five top-operators, two of
them belonging also to the group of the four ones just mentioned
above and three
new ones. We also find that two other "top operators" lead to
anomalous magnetic moment $Z\bar b b$ and $\gamma \bar b b$
couplings, whose observable  first order effects, however, are reduced
by the factor $m_b/m_t$.
Finally our conclusions and an outlook for the future
are given in the last Sect.5.\par

\section{Operators involving third family quarks or bosons}
The complete list of the $dim=6$, $SU(3)_c \times SU(2)\times
U(1)$ invariant operators
involving leptons, quarks, gauge bosons and scalar fields has been
established in ref.\cite{Buchmuller}. Restricting to those operators
involving quarks of the third family only, (\ie\@ either the left
doublet $q_L\equiv (t \ , \ b)_L$ or the right singlets $t_R$, $b_R$),
and bosons and imposing also CP invariance, we obtain the
following set of operators ~classified  in two classes.
In class 1 we put the operators involving at least one $t_R$ field,
while the remaining ones are put in ~class
2. The operators in each class are further ~divided into  two
groups; those containing four quark fields, and those
including only two quark fields: \par

\vspace{0.5cm}
\noindent
{\bf Class 1.}\\
A1) \underline {Four-quark operators}\\
\bqa
\O_{qt} & = & (\bar q_L t_R)(\bar t_R q_L) \ \ \ , \ \\[0.1cm]
\O^{(8)}_{qt} & = & (\bar q_L \overrightarrow\lambda t_R)
(\bar t_R \overrightarrow\lambda q_L) \ \ \ ,\ \\[0.1cm]
\O_{tt} & = & {1\over2}\, (\bar t_R\gamma_{\mu} t_R)
(\bar t_R\gamma^{\mu} t_R) \ \ \ , \ \\[0.1cm]
\O_{tb} & = & (\bar t_R \gamma_{\mu} t_R)
(\bar b_R\gamma^{\mu} b_R) \ \ \ , \ \\[0.1cm]
\O^{(8)}_{tb} & = & (\bar t_R\gamma_{\mu}\overrightarrow\lambda t_R)
(\bar b_R\gamma^{\mu} \overrightarrow\lambda b_R) \ \ \ , \
\eqa
\newpage
\bqa
\O_{qq} & = & (\bar t_R t_L)(\bar b_R b_L) +(\bar t_L t_R)(\bar
b_L b_R)\ \ \nonumber\\
\null & \null & - (\bar t_R b_L)(\bar b_R t_L) - (\bar b_L t_R)(\bar
t_L b_R) \ \ \ , \ \\[0.1cm]
\O^{(8)}_{qq} & = & (\bar t_R \overrightarrow\lambda t_L)
(\bar b_R\overrightarrow\lambda b_L)
+(\bar t_L \overrightarrow\lambda t_R)(\bar b_L
\overrightarrow\lambda  b_R)
\ \nonumber\\
\null & \null &
- (\bar t_R \overrightarrow\lambda b_L)
(\bar b_R \overrightarrow\lambda t_L)
- (\bar b_L \overrightarrow\lambda t_R)(\bar t_L
\overrightarrow\lambda   b_R)
\ \ \  . \
\eqa\\
B1) \underline {Two-quark operators.}\\
\bqa
\O_{t1} & = & (\Phi^{\dagger} \Phi)(\bar q_L t_R\widetilde\Phi +\bar t_R
\widetilde \Phi^{\dagger} q_L) \ \ \ ,\ \\[0.1cm]
\O_{t2} & = & i\,\left [ \Phi^{\dagger} (D_{\mu} \Phi)- (D_{\mu}
\Phi^{\dagger})  \Phi \right ]
(\bar t_R\gamma^{\mu} t_R) \ \ \ ,\\[0.1cm]
\O_{t3} & = & i\,( \widetilde \Phi^{\dagger} D_{\mu} \Phi)
(\bar t_R\gamma^{\mu} b_R)-i\, (D_{\mu} \Phi^{\dagger}  \widetilde\Phi)
(\bar b_R\gamma^{\mu} t_R) \ \ \ ,\\[0.1cm]
\O_{D t} &= & (\bar q_L D_{\mu} t_R)D^{\mu} \widetilde \Phi +
D^{\mu}\widetilde \Phi^{\dagger}
(\ol{D_{\mu}t_R}~ q_L) \ \ \ , \\[0.1cm]
\O_{tW\Phi} & = & (\bar q_L \sigma^{\mu\nu}\overrightarrow \tau
t_R) \widetilde \Phi \cdot
\overrightarrow W_{\mu\nu} + \widetilde \Phi^{\dagger}
(\bar t_R \sigma^{\mu\nu}
\overrightarrow \tau q_L) \cdot \overrightarrow W_{\mu\nu}\ \ \
,\\[0.1cm]
\O_{tB\Phi}& = &(\bar q_L \sigma^{\mu\nu} t_R)\widetilde \Phi
B_{\mu\nu} +\widetilde \Phi^{\dagger}(\bar t_R \sigma^{\mu\nu}
 q_L) B_{\mu\nu} \ \ \ ,\\[0.1cm]
\O_{tG\Phi} & = & \left [ (\bar q_L \sigma^{\mu\nu} \lambda^a t_R)
\widetilde \Phi
 + \widetilde \Phi^{\dagger}(\bar t_R \sigma^{\mu\nu}
\lambda^a q_L)\right ] G_{\mu\nu}^a  \ \ \ .
\eqa

\vspace{0.5cm}
\noindent
{\bf Class 2.}\\
A2) \underline{Four quark ~operators}\\
\bqa
\O^{(1,1)}_{qq} & = &{1\over2}\, (\bar q_L\gamma_{\mu} q_L)
(\bar q_L\gamma^{\mu} q_L) \ \ \ , \\[0.1cm]
\O^{(1,3)}_{qq} & = & {1\over2}\,(\bar
q_L\gamma_{\mu}\overrightarrow\tau q_L) \cdot
(\bar q_L\gamma^{\mu} \overrightarrow\tau q_L) \ \ \ ,\\[0.1cm]
\O^{(8,1)}_{qq} & = & {1\over2}\,(\bar
q_L\gamma_{\mu}\overrightarrow\lambda q_L).
(\bar q_L\gamma^{\mu} \overrightarrow\lambda q_L) \ \ \
,\\[0.1cm]
\O^{(8,3)}_{qq} & = & {1\over2}\,(\bar q_L\gamma_{\mu}\lambda^a\tau^j q_L)
(\bar q_L\gamma^{\mu}\lambda^a\tau^j q_L) \ \ \ , \\[0.1cm]
\O^{(1)}_{bb} & = & {1\over2}\,(\bar b_R\gamma_{\mu} b_R)
(\bar b_R\gamma^{\mu} b_R) \ \ \ ,\\[0.1cm]
\O^{(1)}_{qb} & = & (\bar q_L b_R)(\bar b_R q_L) \ \ \ , \\[0.1cm]
\O^{(8)}_{qb} &= & (\bar q_L\overrightarrow\lambda b_R)\cdot
(\bar b_R\overrightarrow\lambda q_L) \ \ \ .
\eqa\\
\newpage

B2) \underline {Two-quark operators.}\\
\bqa
\O^{(1)}_{\Phi q} &= & i\,(\Phi^{\dagger}  D_{\mu} \Phi)
(\bar q_L\gamma^{\mu} q_L) -i\,(D_{\mu} \Phi^{\dagger}  \Phi)
(\bar q_L\gamma^{\mu} q_L) \ \ \ ,\\[0.1cm]
\O^{(3)}_{\Phi q} &=& i\,\left [( \Phi^{\dagger}
\overrightarrow \tau D_{\mu} \Phi)
-( D_{\mu} \Phi^{\dagger} \overrightarrow \tau \Phi)\right ]
\cdot (\bar q_L\gamma^{\mu}\overrightarrow\tau q_L) \ \ \ ,
\\[0.1cm]
\O_{\Phi b}& = &i\,\left [( \Phi^{\dagger}  D_{\mu} \Phi)
-( D_{\mu} \Phi^{\dagger}\Phi)\right ]
(\bar b_R\gamma^{\mu} b_R) \ \ \ ,\\[0.1cm]
\O_{D b} &=& (\bar q_L D_{\mu} b_R)D^{\mu}\Phi + D^{\mu}\Phi^{\dagger}
(\ol{D_{\mu}b_R} q_L) \ \ \ ,\\[0.1cm]
\O_{bW\Phi} &=& (\bar q_L \sigma^{\mu\nu}\overrightarrow \tau
b_R)\Phi \cdot
\overrightarrow W_{\mu\nu} + \Phi^{\dagger}(\bar b_R \sigma^{\mu\nu}
\overrightarrow \tau q_L)\cdot \overrightarrow W_{\mu\nu}
\ \ \ ,\\[0.1cm]
\O_{bB\Phi} &= & (\bar q_L \sigma^{\mu\nu} b_R)\Phi
B_{\mu\nu} + \Phi^{\dagger}(\bar b_R \sigma^{\mu\nu}
 q_L) B_{\mu\nu} \ \ \ ,\\[0.1cm]
\O_{bG\Phi} &=& (\bar q_L \sigma^{\mu\nu}\lambda^a b_R)\Phi
G_{\mu\nu}^a + \Phi^{\dagger}(\bar b_R \sigma^{\mu\nu}
\lambda^a q_L)G_{\mu\nu}^a \ \ \ ,\
\eqa
where $\lambda^a$ are the eight usual colour matrices.\par
In the preceding ~formulae the usual definitions\\
\bq \Phi=\left( \begin{array}{c}
      \phi^+ \\
{1\over\sqrt2}(v+H+i\phi^0) \end{array} \right) \ \ \ \ , \ \
\eq
\bq
D_{\mu}  =  (\partial_\mu + i~ g\prime\,Y B_\mu +
i~ g \overrightarrow t \cdot \overrightarrow W_\mu ) \ \
\ \  \
\eq
are used where $Y$ is the hypercharge of the field on which the
covariant derivative acts and $\overrightarrow t$ its isospin
matrices.\par

In addition to the above fermionic operators, NP induced by new
heavy degrees of freedom, may also be hiding in purely bosonic
$dim=6$ operators. Provided CP invariance is imposed, this kind
of NP is described by 11 independent $dim=6$ purely bosonic
operators first classified in \cite{Hag}. For completeness
we give them below as \cite{dyn}:\\
\bqa
\overline{\O}_{DW} & =& 2 ~ (D_{\mu} \overrightarrow W^{\mu
\rho}) (D^{\nu} \overrightarrow W_{\nu \rho}) \rangle \ \ \
  , \ \  \\[0.1cm]
\O_{DB} & = & (\partial_{\mu}B_{\nu \rho})(\partial^\mu B^{\nu
\rho}) \ \ \  , \ \ \\[0.1cm]
\O_{BW} & =& \frac{1}{2}~ \Phi^\dagger B_{\mu \nu}
\overrightarrow \tau \cdot \overrightarrow W^{\mu \nu} \Phi
\ \ \  , \ \ \\[0.1cm]
\O_{\Phi 1} & =& (D_\mu \Phi^\dagger \Phi)( \Phi^\dagger
D^\mu \Phi) \ \ \  , \ \ \\[0.1cm]
\O_{\Phi 2} & = & 4 ~ \partial_\mu (\Phi^\dagger \Phi)
\partial^\mu (\Phi^\dagger \Phi ) \ \ \  , \ \ \
\\[0.1cm]
\O_{\Phi 3} & = & 8~ (\Phi^\dagger \Phi) ^3\ \ \  ,
\ \ \\[0.1cm]
\O_W &= & {1\over3!}\left( \overrightarrow{W}^{\ \ \nu}_\mu\times
  \overrightarrow{W}^{\ \ \lambda}_\nu \right) \cdot
  \overrightarrow{W}^{\ \ \mu}_\lambda \ \ \
 ,  \ \ \\[0.1cm]
\widehat{\O}_{UW} & = & \frac{1}{2}\, (\Phi^\dagger \Phi)
\, \overrightarrow W^{\mu\nu} \cdot \overrightarrow W_{\mu\nu} \ \
\  ,  \ \ \\[0.1cm]
\widehat{\O}_{UB} & = & 2~ (\Phi^\dagger \Phi ) B^{\mu\nu} \
B_{\mu\nu} \ \ \  , \ \ \ \\[0.1cm]
\O_{W\Phi} & = & i\, (D_\mu \Phi)^\dagger \overrightarrow \tau
\cdot \overrightarrow W^{\mu \nu} (D_\nu \Phi) \ \ \  , \ \
 \\[0.1cm]
\O_{B\Phi} & = & i\, (D_\mu \Phi)^\dagger B^{\mu \nu} (D_\nu
\Phi)\ \ \  . \ \  \
\eqa\par

As mentioned in the previous section and provided CP invariance
is assumed, NP is described in terms of an effective lagrangian
containing the 14 fermionic operators of the first class given
in (1-14), and the 11 bosonic operators in (31-41).
We then define the effective lagrangian describing the corresponding
residual interactions as
\bq
\L =  \sum_i { f_i \over \Lambda^2}\,\O_i \ \ \ , \
\eq
$\Lambda$ being the NP scale and $f_i$ the dimensionless
coupling of the ~operator $\O_i$.
The observable effects predicted by this lagrangian
will be discussed in the following Sections. At this point we
only note that it is convenient  to remove from $\O_{t1}$
its tree level contribution to $m_t$ by an appropriate
renormalization of the top mass which leads to
\bq
\O_{t1}  \to  (\Phi^{\dagger} \Phi-\frac{v^2}{2})
(\bar q_L t_R\widetilde\Phi +\bar t_R
\widetilde \Phi^{\dagger} q_L) \ \ \ .\
\eq
Similarly, a renormalization of the $W$ and $B$ fields leads to
the ~substitutions
\bqa
\widehat{\O}_{UW} & \to & \frac{v^2}{2}\,\O_{UW}\ \ \ ,\ \\
\widehat{\O}_{UB} & \to & \frac{v^2}{2}\,\O_{UB}\ \ \ , \
\eqa
with the definitions
\bqa
\O_{UW} & \equiv & \frac{1}{v^2}\,(\Phi^\dagger \Phi-\frac{v^2}{2})
\, \overrightarrow W^{\mu\nu} \cdot \overrightarrow W_{\mu\nu} \ \
\  ,  \ \ \\[0.1cm]
\O_{UB} & \equiv & \frac{4}{v^2}\,(\Phi^\dagger
 \Phi-\frac{v^2}{2}) B^{\mu\nu} \,B_{\mu\nu} \ \ \  , \
\eqa
which remove the tree level contributions of these operators to
the $W_\mu$ and $B_\mu$ kinetic energy.\par

\section{Constraints from gauge boson self-energies and light fermions}
The constraints on the couplings of the purely bosonic NP operators
from the available experimental results (mainly) in the light
fermionic sectors  have already been derived in\footnote{Note
Table 1 in the second paper in this Ref.} \cite{Hag}.
For the "non-blind" operators $\ol{\O}_{DW}$, $\O_{DB}$,
$\O_{\Phi 1}$ and $\O_{BW}$, these constraints are so strong
that their relevance
for NP is virtually excluded. Only the "superblind"
operators ($\O_{\Phi 2}$, $\O_{\Phi 3}$), the 5 "blind" operators
($\O_{B\Phi}$, $\O_{W\Phi}$, $\O_{UB}$, $\O_{UW}$, $\O_W$ ) and
of course the above 14 "top" operators have still a chance to
describe an observable NP.
The constraints on the purely bosonic blind operators from  $Z\to b
\bar b$  have also been studied in \cite{RV}, where it has
been found that only $\O_{B\Phi}$ and $\O_{W\Phi}$ are sensitive
to this process, since only these give a $\ln \Lambda^2$
dependent contribution increasing with $m_t$.
 We also note that unitarity considerations have
also been applied to the five "blind" purely bosonic
operators. They led to the conclusion that
"unitarity" is as effective in constraining the "blind"
couplings, as are present LEP1 measurements \cite{unitarity}.
\par

In this Section we give the constraints for the
"top" operators of our first Class.
These operators contribute to the light fermion
processes only at the 1-loop level, giving
universal oblique corrections to the gauge boson self-energies.
In general, the relevant diagrams  have the same topology
as the SM ones; \ie\@ $t\bar t$ loops
for neutral currents and $t \bar b$ loops for charged currents  (in
some cases tadpoles generated by 4-leg couplings may also appear). In
the SM, these diagrams produce the well-known strong $m^2_t$ contribution to
$\Delta\rho$. For the top operators listed in (1-14), contributions
having a different $m_t$ dependence may be generated.
In the calculation, we only
keep the divergent part of the leading $m_t$
contribution. This is required for consistency with our
effective lagrangian
approach, where we restrict to $dim=6$ operators only.\par

Only four of the above "top" operators give a non vanishing
NP contribution to either the $\epsilon_1$ or $\epsilon_3$
parameters  conventionally defined in \cite{AB,Peskin}.
All other "top" operators give no contribution to
$\epsilon_{1,\,3}$ and none of the operators  contributes
to $\epsilon_2$.
Thus, defining $L \equiv \ln{\Lambda^2/M_Z^2}$, the only non vanishing
results are:
\bq
\epsilon^{(NP)}_1(t2) = -{3m^2_t\over4\pi^2\Lambda^2}f_{t2}L
= -0.011f_{t2}\ \ \
\eq
from $\O_{t2}$,
\bq
\epsilon^{(NP)}_1(Dt) =
-{3gm^3_t\over16\pi^2\sqrt2 M_W\Lambda^2}f_{Dt}L = -0.0028f_{Dt}
\ \ \
\eq
for $\O_{Dt}$,
\bq
\epsilon^{(NP)}_3(tW\Phi) =
-{5M_W m_t\over4\pi^2\sqrt2\Lambda^2}f_{tW\Phi}L
= -0.0060f_{tW\Phi} \ \ \
\eq
for $\O_{tW\Phi}$ and
\bq
\epsilon^{(NP)}_3(tB\Phi) =
-{3\cw M_W m_t\over4 \pi^2\sqrt2 \sw \Lambda^2}f_{tB\Phi}L
= -0.0066f_{tB\Phi} \ \ \
\eq
for $\O_{tB\Phi}$.
For the numerical results in (48-51) we have used $m_t=175 GeV$
and $\Lambda=1TeV$, while $\swd$ has been identified with
$s_0^2 \simeq 0.231$  defined
by $ s_0c_0=\pi\alpha (M_Z)/(\sqrt2 M^2_Z
G_\mu)$ and describing the Weinberg angle including QED
corrections only  \cite{AB}.\par

The present experimental knowledge from LEP1 and SLC is
summarized \eg\@ in \cite{lep1}, where it
is found that
\bq
-3.2\times10^{-3} \lsim \epsilon^{(NP)}_1 \lsim +3.2\times10^{-3}\ \ \ ,
 \eq
\bq
-3.8 \times10^{-3}\lsim \epsilon^{(NP)}_3 \lsim +1.8\times10^{-3} \ \ \ ,
\eq
provided $m_t$, $m_H$ are ~allowed to vary in the range
$160 \lsim m_t \lsim 190 GeV$ and $65 GeV\lsim m_H \lsim
1 TeV $. Comparing (52,53) with (48-51) one then gets
\bq  -0.3 \lsim f_{t2} \lsim +0.3 \ \ \ ,\eq
\bq  -1.1 \lsim f_{Dt} \lsim +1.1 \ \ \ ,\eq
\bq  -0.27 \lsim f_{tW\Phi} \lsim +0.47 \ \ \ , \eq
\bq  -0.27 \lsim f_{tB\Phi} \lsim +0.43 \ \ \ , \eq
provided that each operator is considered separately, and that no
{}~cancellations among the contributions from different operators are
taken into account.\par

\section{Constraints from the $b\bar b$ observables}
At 1-loop the top quark operators also affect  the $Zb \bar b$
and $\gamma b \bar b$ couplings. In the
SM case, the top and goldstone
(in the $R_\xi$ gauge) exchange diagrams produce the well-known
strong $m^2_t$ contribution. With our set of top operators one
generates several new $m_t$ dependent contributions.
Again, for each operator, we only retain the leading $m_t$
and $\ln\Lambda^2$
dependent contributions, and neglect
quantities proportional to $m_b/M_Z$. Non-vanishing effects now
arise only from the five four-quark operators $\O_{qt}$, $\O^{(8)}_{qt}$,
$\O_{tb}$, $\O_{qq}$, $\O^{(8)}_{qq}$, and from the two two-quark
operators $\O_{t2}$ and $\O_{Dt}$.
These operators give 3 different types of anomalous
contributions. Namely, vector
and axial vector couplings for  $Zb\bar b$, and anomalous magnetic
moment couplings for both $Zb\bar b$ and $\gamma b\bar b$.
We normalize the vector and axial $Zb\bar b$ vertex
(S-matrix elements) as\footnote{Note that charge conservation prohibits the
appearance of anomalous vector and axial couplings for $\gamma$.}
\bq
({-ie\over2s_Wc_W})\gamma^{\mu}[g^Z_{Vb} + \delta g^Z_{Vb}
-\gamma^5(g^Z_{Ab} + \delta g^Z_{Ab})] \ \ \ ,
\eq
with $g^Z_{Vb}=(-1/2+2\swd/3)$ , $g^Z_{Ab}=-1/2$,
and the anomalous $Z$ and $\gamma$ magnetic moment couplings by
\bq
{e\over2\sw \cw m_t}(\sigma^{\mu\nu}q_{\nu})\,\delta \kappa^Z \ \ \ ,
\eq
\bq
{e\over m_t}(\sigma^{\mu\nu}q_{\nu})\,\delta \kappa^{\gamma} \ \ \ .
\eq\par

Turning now to the results, we start from the remark that
the operators
$\O_{qt}$, $\O^{(8)}_{qt}$, $\O_{t2}$ and  $\O_{Dt}$ give purely
left-handed contributions to the anomalous $Zb\bar b$ coupling.
These are written  as
\bq
\delta g^Z_{Vb} = \delta g^Z_{Ab}= \frac{L}{32\pi^2
\Lambda^2}\cdot
\Big [\, (f_{qt}+\frac{16f^{(8)}_{qt}}{3}-f_{t2})\,m^2_t
 + {5gf_{Dt}m^3_t\over2 \sqrt2 M_W} \,\Big ] \ .
\eq
On the contrary, the operator
$\O_{tb}$ generates  a pure right-handed
NP contribution to $Zb\bar b$, which is given by
\bq
\delta g^Z_{Vb} = -\delta g^Z_{Ab} =
-{3f_{tb}m^2_t\over16\pi^2\Lambda^2}L \ \ \ . \
 \eq
Finally, $\O_{qq}$ and
$\O^{(8)}_{qq}$  generate only anomalous magnetic
moment-type couplings for both, $Z$ and $\gamma$. Using the
definitions (59,60) we find
\bqa
\delta \kappa^Z & = &-(f_{qq}+
{16\over3}f^{(8)}_{qq}){m^2_t(1-8\swd/3)
\over32\pi^2\Lambda^2}L\ \ \ ,\\
\delta \kappa^{\gamma} &= & -(f_{qq}+
{16\over3}f^{(8)}_{qq}){2m^2_t
\over48\pi^2\Lambda^2}L \ \ \ .
\eqa\par

The interesting thing about these anomalous magnetic couplings
is that they have nothing to do with the b-quark mass $m_b$;
\ie\@ they can exist even if $m_b$ vanishes. Their contribution to
observable effects is however, to first order, proportional to
$m_b/m_t$. This is easily understood because first order
contributions could only arise from ~interference with the SM
amplitude, which, ~being vector or axial, leads to
$(b,~ \bar b)$ pairs with ~opposite helicities, while the
magnetic interactions induced by
$\O_{qq}$ or $\O^{(8)}_{qq}$ want to give
to $(b,~ \bar b)$ the same helicity. Thus, in the $m_b \to 0$
limit there is no ~interference.
We should also remark that the treatment of $\O_{qq}$ and
$\O^{(8)}_{qq}$ to first order only is consistent with our
approximation to
neglect $dim=8$ operators, which will ~inevitably arise in the
divergent part of loops involving two $dim=6$ "top"
operators.\par

We conclude therefore that seven of the 14 "top"
operators give NP contributions to $Z\to b \bar b$.
These contributions, determined by (58 - 64), modify the partial width
$\Gamma (Z\to b \bar b) \equiv \Gamma_b$  and
the "longitudinally polarized forward-backward asymmetry" $A_b$
defined at the $Z$ peak by
\bqa
A_b  & \equiv & \frac{\sigma (e^-_L\to b_F) - \sigma ( e^-_L \to b_B) +\sigma
(e^-_R\to b_B) - \sigma ( e^-_R \to b_F)}
{\sigma (e^-_L\to b_F) + \sigma ( e^-_L \to b_B)+ \sigma
(e^-_R\to b_B) + \sigma ( e^-_R \to b_F)}
\nonumber \ \ \ \\
\null & = &\frac{\sigma (e^-_L\to b_F) - \sigma ( e^-_L \to b_B)}
    {\sigma (e^-_L\to b_F) + \sigma ( e^-_L \to b_B)}
   = \frac{\sigma (e^-_R\to b_B) - \sigma ( e^-_R \to b_F)}
    {\sigma (e^-_R\to b_B) + \sigma ( e^-_R \to b_F)}
\ \ \ , \
\eqa
where the second line in (65) just follows by rotating the $Z$ spin by
$180^0$ ~around an axis perpendicular to the beam direction.
In \cite{CRV}, it has been shown that from these quantities one
can measure two model independent parameters which are sensitive
to the NP ~considered in the present work, namely
\bq
{\Gamma_b \over \Gamma_s} \equiv 1 + \delta_{bv}
\ \ \ ,\eq
\bq {A_b\over A_s} \equiv 1+\eta_b \ \ \ . \eq\par

The New Physics (NP) contributions to these
parameters are
\bqa
\delta^{(NP)}_{bv}& =& -{4\over 1+v^2_d}\,[v_d\delta g^Z_{Vb}
+\delta g^Z_{Ab}+3v_d{m_b\over m_t}\delta \kappa^Z] \ \ \ , \\[0.1cm]
\eta^{(NP)}_{b} &= &-{2(1-v^2_d)\over v_d(1+v^2_d)}\,[\delta g^Z_{Vb}
-v_d\delta g^Z_{Ab}+] -{4(1-2v^2_d)\over v_d(1+v^2_d)}{m_b\over m_t}
\delta \kappa^Z\,\ \ \ ,
\eqa
where $v_d=1-\frac{4}{3}s_0^2$, and $ s_0^2 \simeq 0.231 $ has already
been defined  immediately after (51). Using (61-64) we thus find
\bq
\delta^{(NP)}_{bv}= -0.0021(f_{qt}+\frac{16}{3}f^{(8)}_{qt}-f_{t2})
-0.0048f_{Dt}-0.0023f_{tb}
+0.17\times10^{-4}(f_{qq}+\frac{16}{3}f^{(8)}_{qq}) ,
\eq
\bq \eta^{(NP)}_{b} =  -0.00014(f_{qt}+\frac{16}{3}f^{(8)}_{qt}
-f_{t2})
-0.00036f_{Dt}+0.0046f_{tb}
+5.4\times10^{-7}(f_{qq}+\frac{16}{3}f^{(8)}_{qq})  ,
\eq
where the same input parameters as in the preceding section have
been used.\par

It is worth noting from (66, 67) that the parameters
$\delta_{bV}$ and $\eta_b$ are useful
for any kind of coupling, while the parameter
$\epsilon_b$ defined in \cite{AB} applies only
to the pure left-handed case for which it is given by
$ \epsilon_b=-2\delta g^Z_{Vb}=-2\delta g^Z_{Ab}$.
Using (68, 69) we also notice for the NP contribution
that the sign (and magnitude) of the ratio
$\eta^{(NP)}_b/\delta^{(NP)}_{bv}$ discriminates between
the purely left handed
or the magnetic anomalous contribution on the one side, and the
purely right handed one induced by $\O_{tb}$. Indeed we find
\bq
\eta^{(NP)}_b/\delta^{(NP)}_{bv} ~=~ \frac{(1-v_d)^2}{2v_d} ~=~0.068 ~>
{}~ 0 \ \
\eq
for $(\O_{qt},\,\O^{(8)}_{qt},\,\O_{Dt},\,\O_{t2})$,
and
\bq
\eta^{(NP)}_b/\delta^{(NP)}_{bv} ~=~ \frac{(1-2v_d^2)}{3v_d^2}~
=~0.03 ~> ~ 0 \ \
\eq
for $(\O_{qq},\,\O^{(8)}_{qq})$, while the $\O_{tb}$ case gives
\bq
\eta^{(NP)}_b/\delta^{(NP)}_{bv} ~=~ -\,\frac{(1+v_d)^2}{2v_d} ~
=~-2.068 ~< ~ 0 \ \ .\ \eq
This numerical difference between the predictions (74) and
(72,73) could be essential in the search for the $\O_{tb}$ operator
at SLC.\par

The results presently available on $\Gamma_b$ alone from
LEP \cite{lep1,lep1b}
and SLC \cite{slc}, would lead to a  difference between the
experimental findings and the SM prediction:
\bq
\delta^{(NP)}_{bV}=(+1.93\pm1.08)\times10^{-2} \ \ \ .
\eq
By comparing this with (69)
one obtains the following one-standard deviation numerical constraints on
the coupling constants of the
contributing seven "top" operators,
 taken one by one:
\bq  -15 \lsim f_{qt} \lsim  -4 \ \ \ , \eq
\bq  -3 \lsim f^{(8)}_{qt} \lsim  -0.7  \ \ \ ,\eq
\bq  -6 \lsim f_{Dt} \lsim  -2  \ \ \ , \eq
\bq  +4 \lsim f_{t2} \lsim  +15  \ \ \ ,\eq
\bq  -14 \lsim f_{tb} \lsim  -4  \ \ \ , \eq
\bq 0.5\times10^{+3} \lsim f_{qq} \lsim 2\times10^{+3}\ , \eq
\bq 10^{+2} \lsim f^{(8)}_{qq} \lsim 4\times10^{+2} \ .
\eq \par

The very loose limit on $f_{qq}$ and $f^{(8)}_{qq}$ is due to the
presence of the $m_b/m_t$ factor in front of the magnetic coupling
$\delta \kappa^Z$ in (68,69). It corresponds to a $\delta \kappa^Z$
value of the order of 0.1. One may wonder whether it could be possible
to measure separately the magnetic $\gamma b \bar b$ and $Zb \bar b$
couplings by performing measurements outside
the Z peak.
The differential cross section for the process $e^+e^- \to b \bar b$
going through photon and Z exchange,
calculated at the tree level and neglecting for consistency
quadratic terms in $(\delta \kappa^\gamma)$ and $(\delta \kappa^Z)$, is
given by
\bqa
\frac{d\sigma}{d\Omega} & =& \frac{\alpha^2 \beta_b}{4s}
\Bigg \{ Q_b^2 \left (1 +\beta_b^2
\cos^2 \theta + \frac{4 m_b^2}{s} \right ) +
\frac{8m_b}{m_t} Q_b \delta \kappa^\gamma \nonumber \\
\null & \null & +~ \frac{s^2}{16s^4_Wc^4_W|D_Z|^2} \Bigg [
(g_{Ve}^2+g_{Ae}^2)
 \left \{ (g_{Vb}^2+g_{Ab}^2)(1+\beta_b^2 \cos^2\theta )+
(g_{Vb}^2-g_{Ab}^2)
\frac{4 m_b^2}{s}  \right \}  \nonumber \\
\null & \null & + ~8 g_{Ve} g_{Ae} g_{Vb} g_{Ab} \beta_b \cos
\theta   + 8\delta \kappa^Z ~\frac{m_b}{m_t}
\{ (g_{Ve}^2+g_{Ae}^2)g_{Vb} +2 g_{Ve} g_{Ae} g_{Ab}
\beta_b \cos \theta \}
\Bigg ] \nonumber \\
\null & \null & -~\frac{ s (s -M_Z^2)}{2s^2_Wc^2_W|D_Z|^2}
\Bigg [Q_b g_{Ve} g_{Vb} \left (1+\beta_b^2
\cos^2 \theta + \frac{4 m_b^2}{s} \right ) \nonumber \\
\null & \null & +~
2 Q_b g_{Ae} g_{Ab} \beta_b \cos \theta
+ 4\frac{m_b}{m_t} (g_{Ve}[Q_b \delta \kappa^Z +
g_{Vb}\delta \kappa^\gamma] +g_{Ae} g_{Ab} \beta_b
\delta \kappa^\gamma\,
\cos \theta ) \Bigg ] \Bigg \} ,
\eqa
where for $Q_f$ is the fermion charge,
\bq
g_{Vf} = t_f^{(3)}-2Q_f\swd \ \ \ , \ \ \
g_{Af}= t_f^{(3)}\ \ \ , \
\eq
$\beta_b=\sqrt{1-4m_b^2/s}$ is the $b$ quark velocity and
$|D_Z|^2=(s-M_Z)^2+ M_Z^2\Gamma_Z^2 $.\par

We see from (83), that an accuracy of one percent below the Z
peak would allow the determination of
$\delta \kappa^\gamma$ at the level of 0.1 . This would mean
roughly  the same
sensitivity to $f_{qq}$ and $f^{(8)}_{qq}$ as from Z peak
experiments. Anomalous magnetic moment interactions have also
been studied in \cite{Zerwas}. \par

\section{Conclusions}
In this paper we have studied some of the New Physics signatures
expected in the case where all the new degrees of
freedom are too heavy to be directly produced at the Colliders
in the ~foreseeable future. In such a case NP is ~predominantly described by
$dim=6$ operators involving only standard model particles,
including the usual Higgs doublet. Motivated by the overall
picture implied by the amazing success of the SM in ~explaining the
present precision measurements, we are led to a set of 39 $SU(3)_c\times
SU(2)\times U(1)$ gauge invariant and CP symmetric operators.
Eleven of these operators are purely bosonic and have been
studied before, while the remaining 28 involve
in addition quark fields of the third family.
Among these 28 operators, there are 14 where the $t_R$ field
appears, at least once.
 The motivation for singling out the quarks of
the third family is ~supplied by the large top mass, which
indicates a strong "affinity" of these quarks to the Higgs sector.
If we believe that a next possible step in particle
physics is that of  understanding  the
{}~spontaneous breaking mechanism, then a good way
to find some kind of new physics is that of looking
whether any of these operators ~acquires an observable strength.
In this respect it looks as if the $t_R$ involving operators, as
well as the purely bosonic ones, are more likely to be enhanced
by whatever NP is hidden in the scalar sector.  \par

The above 14 "top" operators should best be studied
through their effects in top production at the future Colliders.
Before doing this, though, we need to study what kind of hints
on the expected ~strength of the various operators may be obtained
from LEP1 and SLC. Thus in the present paper we have studied their
effects  on the gauge boson self energies and
the $Z \to b \bar b$ decay. It turns out that five of these
operators, namely $\O_{tt},\,\O^{(8)}_{tb},\,\O_{t1},\,\O_{t3}$
and $\O_{tG\Phi}$, give no contribution to these quantities.
Thus, present experimental knowledge provides no information on
them. On the other hand,
the remaining nine operators give non vanishing contributions to
at least one of $\epsilon_1,\,\epsilon_3$ and the $Z\to b \bar
b$ parameters $\eta^{(NP)}_b$ and $\delta^{(NP)}_{bv}$. The
results are summarized in Table 1, where the blanks indicate no
contribution from the corresponding operator.
\noindent
\begin{center}
\begin{tabular}{|c|c|c|c|c|} \hline
\multicolumn{5}{|c|}{Table 1: Contributions of "top" operators
to $Z$ peak physics.}\\[.1cm] \hline
\multicolumn{1}{|c|}{Operator} &
  \multicolumn{1}{|c|}{$\epsilon^{(NP)}_1$} &
   \multicolumn{1}{|c|}{$\epsilon^{(NP)}_3$ } &
     \multicolumn{1}{|c|}{$\delta^{(NP)}_{bv}$} &
       \multicolumn{1}{|c|}{$\eta^{(NP)}_b /\delta^{(NP)}_{bv}$}
          \\[0.1cm] \hline
  $\O_{qt}$ & \null & \null & $-2.1\times 10^{-3}f_{qt}$ & $0.068$ \\[0.1cm]
  $\O^{(8)}_{qt}$ & \null & \null & $-1.1\times
         10^{-2}f^{(8)}_{qt}$ & 0.068 \\[0.1cm]
  $\O_{t2}$ & $-1.1 \times 10^{-2}f_{t2} $ & \null & $2.1\times
       10^{-3}f_{t2}$  & $0.068$ \\[0.1cm]
  $\O_{Dt}$ &$-2.8 \times 10^{-3} f_{Dt}$ & \null &$-4.8\times
      10^{-3}f_{Dt}$   & $0.068$ \\[0.1cm]
  $\O_{qq}$ & \null & \null & $1.7\times 10^{-5}f_{qq}$  & $0.03$ \\[0.1cm]
  $\O^{(8)}_{qq}$ & \null & \null & $9.1\times
         10^{-5}f^{(8)}_{qq}$ & $0.03$ \\[0.1cm]
$\O_{tb}$ & \null & \null & $-2.3 \times 10^{-3}f_{tb}$ &
$-2.068$ \\ [0.1cm]
$\O_{tW\Phi}$ & \null & $-6.0\times 10^{-3}f_{tW\Phi}$ & \null &
\null \\[0.1cm]
  $\O_{tB\Phi}$ & \null & $-6.6\times 10^{-3}f_{tB\Phi}$ & \null &
  \null \\ \hline
 \end{tabular}
\end{center}
\noindent
It should be noted that none of these operators contribute to
$\epsilon_2$.\par

The most interesting  result in Table 1 is given by its last column
which indicates that the ratio $\eta^{(NP)}_b
/\delta^{(NP)}_{bv}$ provides a very strong signature
for discriminating between the left-handed, right-handed and
the anomalous magnetic $Zb\bar b$ vertex. Note that if a single
operator dominates,  the ratio $\eta^{(NP)}_b
/\delta^{(NP)}_{bv}$ is independent of the magnitude of its
coupling and depends
only on the nature of the induced $Zb\bar b$ vertex.\par

It should be stressed that the large and negative
$\eta^{(NP)}_b /\delta^{(NP)}_{bv}$ ratio would be a rather peculiar
signature of the $\O_{tb}$ operator. In practice, it would predict
a two percent (negative) effect in $\eta^{(NP)}_b$ for a one percent
positive effect in $\delta^{(NP)}_{bv}$. This should be detectable at SLC
at their expected final accuracy. Note that this
effect would be
of opposite sign (and larger in magnitude) than  the
correponding
prediction for the remaining operators $\O_{qt}$, $\O^{(8)}_{qt}$,
$\O_{t2}$, $\O_{Dt}$, $\O_{qq}$, $\O^{(8)}_{qq}$ that
contribute here.
Note also that two of these operators,
namesly $\O_{t2}$ and $\O_{Dt}$ are
(qualitatively at least) disfavoured by our analysis from the apparent
inconsistency between their effects on $\epsilon^{(NP)}_1$ indicated in
(54,55) and on $\delta^{(NP)}_{bv}$ shown in (78,79).\par
 Finally, it is
 is more spectacular to remark, that the predicted ratio
$\eta^{(NP)}_b/\delta^{(NP)}
_{bv}$ and the magnitude of $\eta^{(NP)}_b$ for the $\O_{tb}$ operator
 would be orthogonal to the  expectations for the minimal
  supersymmetric SM. Here, in fact, the trend would be that of
\underline{positive} $\eta^{(NP)}_b$ (of order one percent) for
positive $\delta^{(NP)}_{bv}$. However, this prediction would be
necessarily acompanied by the discovery of suitably light
supersymmetric particles, like \eg\@ a light chargino and/or a
light neutral Higgs.

\underline{Acknowledgements}: GJG would like to thank the
Montpellier particle theory group for the very warm hospitality
he ~enjoyed there.\par


\newpage



\begin{thebibliography}{99}

\bibitem{lep1} See e.g. the talk given by G. Altarelli at the
Rome Conference on "Phenomenology of Unification from Present to
Future"; Proc. of the EPS Conference on High
Energy Physics, Marseille, France, 1993 CERN-TH.7319/94,
CERN-TH.7045/93 July 1993.
G. Quast, {\it{ibid.}}; J. Lefran\c{c}ois
{\it{ibid.}}; D. Schaile, CERN-PPE/93-213 (1993); LEP Electroweak
Working Group, CERN-PPE/93-157 (1993); The LEP Collaborations, ALEPH,
DELPHI, L3, OPAL and the LEP Electroweak Working Group, CERN/PPE/94-187
; G.Altarelli, CERN-TH.7464/94.
%
\bibitem{lep1b} D. Schaile, presented at the
27th Int. Conf. on High Energy Phys., Glasgow, (1994). J.Erler and
P. Langacker, UPR-0632T (1994).
\bibitem{slc}SLD Collaboration, K.Abe et al.,\prl{73}{1994}{25}.
%
\bibitem{BMT} G.J. Gounaris et al, in Proc. of the Workshop on $e^+e^-$
 Collisions a  500
GeV:The Physics Potential, DESY 92-123B(1992), p.735,ed. P. Zerwas;
 M. Bilenky, J.L. Kneur, F.M. Renard and
D.~Schildknecht, \np{B419}{1994}{240}.
 M. Bilenky, J.L. Kneur, F.M. Renard and D.
{}~Schildknecht,\np{B409}{1993}{22}.
%
\bibitem{Hag}
K.Hagiwara et al, \pl{B283}{1992}{353};\pr{D48}{1993}{2182}.
%
\bibitem{DeR} A. De R\'{u}jula \etal\@, \np{B384}{1992}{3}.
%
\bibitem{topfnal} CDF Collaboration, (F. Abe \etal\@
\prl{73}{1994}{225}, FNAL-PUB-94/097-E;
S.Abachi et al, D0 Collaboration, \prl{72}{2138}{1994}.
%
\bibitem{Georgi} H. Georgi \np{B361}{1991}{339}.
%
\bibitem{RV}F.M. Renard and C. Verzegnassi, CERN-TH.7376/94 (1994).
\bibitem{Peccei}R.D.Peccei, S.Peris and X.Zhang, \np{B349}{1990}{305};
E.Malkawi and C.P.Yuan, MSUHEP-94/06.
\bibitem{dyn}G.J.Gounaris, F.M.Renard and G.Tsirigoti,
\pl{B338}{1994}{51}.
\bibitem{CRV} D. Comelli, F.M. Renard and C. Verzegnassi, \pr
{D50}{1994}{3076}.
\bibitem{Buchmuller} W. Buchm\"{u}ller and D. Wyler,
\np{B268}{1986}{621}; C.J.C. Burgess and H.J. Schnitzer,
\np{B228}{1983}{454}; C.N. Leung, S.T. Love and S. Rao
\zp{C31}{1986}{433}.
\bibitem{unitarity} G.J. Gounaris, J. Layssac and F.M. Renard,
\pl{B332}{1994}{146}; G.J. Gounaris, J. Layssac, J.E. Paschalis
and F.M. Renard, preprint PM/94-28, to appear in Zeit.f. Physik.
\bibitem{AB} G.Altarelli, R.Barbieri, F.Caravaglios, \pl {B314} {357}
{1993}.
\bibitem{Peskin} M.Peskin and T.Takeuchi, \pr {D46} {381} {1992}.
\bibitem{Zerwas} G. K\"{o}pp, D.Schaile, M. Spira and P.M.
Zerwas, DESY 94-148.

















\end{thebibliography}
\end{document}